\documentclass[pra,aps,showpacs,superscriptaddress,twocolumn,floatfix]{revtex4}
\usepackage{graphicx}
\usepackage[ansinew]{inputenc}
\usepackage{array}
\usepackage{color}
\usepackage{amsmath}
\usepackage{amsxtra}
\usepackage{amstext}
\usepackage{amssymb}
\usepackage{latexsym}
\newcommand{\bea}{\begin{eqnarray}}
\newcommand{\eea}{\end{eqnarray}}
\newcommand{\be}{\begin{equation}}
\newcommand{\ee}{\end{equation}}

\begin{document}

\title{Quantum back-action in spinor condensate magnetometry} 
\author{S. K. Steinke}
\affiliation{B2 Institute, Department of Physics and College of Optical Sciences\\The University of Arizona, Tucson, Arizona, 85721.}

\author{S. Singh}
\affiliation{ITAMP, Harvard-Smithsonian Center for Astrophysics, Cambridge, Massachusetts 02138.}

\author{P. Meystre}
\affiliation{B2 Institute, Department of Physics and College of Optical Sciences\\The University of Arizona, Tucson, Arizona, 85721.}

\author{K. C. Schwab}
\affiliation{Applied Physics, California Institute of Technology, MC 128-95, Pasadena, California 91125.}

\author{M. Vengalattore}
\affiliation{Laboratory of Atomic and Solid State Physics, Cornell University, Ithaca, New York, 14853.}

\date{\today}
\pacs{07.55.Ge, 42.50.Lc, 03.75.Gg}
\begin{abstract}
We provide a theoretical treatment of the quantum backaction of Larmor frequency measurements on a spinor Bose-Einstein condensate  by an off-resonant light field. Two main results are presented; the first is a ``quantum jump'' operator description that reflects the abrupt change in the spin state of the atoms when a single photon is counted at a photodiode. The second is the derivation of a conditional stochastic master equation relating the evolution of the condensate density matrix to the measurement record. We  comment on applications of this formalism to metrology and many-body studies. 
\end{abstract}

%\pacs{42.50.Lc, 42.50.Wk, 42.79.Gn, 07.10.Cm}

\maketitle
Atomic vapor magnetometers of spin-polarized alkali atoms are among the most sensitive field sensors demonstrated to date \cite{budker}. These magnetometers, based on the optical detection of Larmor precession, have demonstrated field sensitivities in the attoTesla/Hz$^{1/2}$ regime \cite{dang2010}. The use of optically trapped ultracold atoms as the sensing medium holds promise for magnetic microscopy at high spatial resolution as well as for significant improvements in field sensitivity via entanglement-assisted techniques \cite{petersen2005, auzinsh04,wasilewski2010}. Spinor Bose-Einstein condensates are particularly suited to field sensing applications due to their low spin relaxation rates and absence of density-dependent collision shifts \cite{mvmag}. 

The detection of Larmor precession and the subsequent estimate of the magnetic field relies on the dispersive interaction between the collective atomic spin and the optical field, followed by a quantum-limited measurement of the light. This interaction entangles the optical and atomic degrees of freedom. Measuring the light breaks this entanglement and must therefore cause a backaction on the atomic spins. The ultimate sensitivity of such atomic magnetometers is governed by the interplay between the projection noise of the atomic spins, photon shot noise and the quantum backaction due to the measurement of the optical field. Here, we provide a theoretical treatment of this backaction by means of a conditional stochastic master equation that relates the condensate evolution to the optical measurement record. One of the main results that distinguishes this work from similar studies, e.g. \cite{Thomsen2002}, is the derivation of a quantum jump operator reflecting the backaction on the atoms induced by the detection of a single photon.

{\em Model:}
We consider a magnetometer consisting of a spinor Bose-Einstein condensate of $n \gg 1$ spin-1 bosons trapped in their motional ground state.
Our formalism is well suited to both spinor gases with ferromagnetic (e.g. $^{87}$Rb) and those with polar (e.g. $^{23}$Na) ground states. 
 %This could be for instance a rubidium condensate with a ferromagnetic ground state or a sodium condensate with an anti-ferromagnetic ground state. 
We assume in this paper that the spatial extent of the gas is small enough that the single-mode approximation (SMA) is valid. We also note that our results can be readily adapted to systems with different internal state spaces or expanded to incorporate spatial variations in the atomic spin field.

The atoms are prepared initially with high fidelity in the spin state, 
\be
|\psi\rangle = c_+|+1\rangle+c_0|0\rangle+c_-|-1\rangle
\ee
in e.g. the $z$ basis, with the normalization $\sum_i |c_i|^2=1$.
%The full $n$-particle state is then given by
%\begin{eqnarray}
%\label{longinitcond} %%%%
%|\Psi_n\rangle &= &\hat a_\psi^{\dag n}|0\rangle/\sqrt{n!}\\
%&=& \sum_{j_++j_0+j_-=n}\sqrt{\frac{n!}{j_+!j_0!j_-!}} c_+^{j_+}c_0^{j_0}c_-^{j_-} |j_+,j_0,j_-\rangle,%\nonumber
%\end{eqnarray}

Before proceeding further, we introduce a ``vector'' notation over the spin indices for clarity and compactness, with the definitions
\bea
\mathbf v & = & \left ( v_+, v_0, v_- \right ),\\
|\mathbf v| & = & v_++v_0+v_-,\\
\mathbf v ! & = & v_+!v_0!v_-!,\\
\mathbf v^{\mathbf w} & = & v_+^{w_+}v_0^{w_0}v_-^{w_-},
\eea
and the dot product defined in the usual way. For later use, we also define probabilities
\be
p_i = |c_i|^2,
\ee
where the state normalization implies that $|\mathbf p| = 1$. With this notation, the full $n$-particle state can be compactly expressed as
\be
\label{shortinitcond}%%%%
|\Psi_n\rangle = \hat a_\psi^{\dag n}|0\rangle/\sqrt{n!} = \sum_{|\mathbf j|=n} \sqrt{\frac{n!}{\mathbf j!}} \mathbf c^{\mathbf j} |\mathbf j\rangle,
\ee
where $\hat{a}_\psi^\dag = {\mathbf c}\cdot \hat{\mathbf a}^\dag$ is the Schr\"{o}dinger field creation operator for state $|\Psi\rangle$, $|0\rangle$ is the vacuum, and $|{\mathbf j}\rangle$ is the state with $j_i$ atoms in spin eigenstate $i$.
%The characterization of the internal state of the condensate is carried out via a homodyne detection scheme where the condensate sits in one branch of a Mach-Zehnder  interferometer. The path lengths of both arms are adjusted so that the difference between the measured outgoing photocurrents is proportional to the relative phase shift of the light in the ``sample'' branch -- the branch containing the condensate. A coherent, polarized short optical pulse of dimensionless amplitude $A_0$ (taken to be real without loss of generality) and mean photon number $|A_0|^2$ is sent into the interferometer through a 50-50 beam splitter. We assume that it is near resonance with a specific electronic resonance yet sufficiently far detuned that the upper level can be adiabatically eliminated and that absorption is negligible -- although absorptive effects could straightforwardly be added to the model. After the light passes through the two branches and recombines at a second 50/50 beam splitter, the output field intensities at the two exit ports are recorded. 

The spin state of the condensate is optically detected either by measuring the phase imprinted onto a short, coherent pulse of the light field by the condensate as in phase-contrast imaging \cite{carusotto2004,higbie}, or by monitoring the polarization of the light field as in polarization spectroscopy (see, for example \cite{budker2002}). Without loss of generality, we assume the former method via balanced homodyne detection.

By introducing a general vector of coupling strengths $\mathbf g$, we can 
describe the light-atom interaction via the Hamiltonian
\be
\label{compactH}%%%%
H_{I}=\hbar \mathbf g\cdot\hat{\mathbf N} \hat b^\dag \hat b.
\ee
where $N_i = \hat{a}_i^\dagger \hat{a}_i$ and $\hat b$ is the light field annihilation operator. To derive this Hamiltonian, we have assumed that the probe field is detuned sufficiently far from the atomic resonance so that the excited atomic states can be adiabatically eliminated\cite{supplemental}. The effect of this Hamiltonian is to imprint a phase rotation on the optical field that contains information about the spin state of the atoms. The optical transit time is taken to be short enough that other contributions to the condensate's evolution may be neglected for now. The quantum state of the combined atomic spin and two output optical fields after passing through the final beam splitter, just before measurement, is therefore
\be
\label{premeasure}%%%%
\sum_{|\mathbf j|=n}\sqrt{\frac{n!}{\mathbf j!}} \mathbf c^{\mathbf j} |\mathbf j\rangle\otimes|\frac{A_0}{2}(1+i e^{-i\mathbf G\cdot\mathbf j})\rangle\otimes|\frac{A_0}{2}(1-ie^{-i\mathbf G \cdot \mathbf j})\rangle,
\ee
where $A_0$ is the dimensionless amplitude of the coherent optical pulse (i.e., the mean photon number is $|A_0|^2$), $\tau$ is the light-matter interaction time and  $\mathbf G = \tau\mathbf g.$
The probe beam's contribution to the output fields has been phase rotated through the unitary evolution given by the Hamlitonian \eqref{compactH}. Importantly, unless the atoms are in the unlikely configuration of a Fock state of the spin field, this interaction entangles the optical field with the collective spin state of the atoms \cite{kuzmich1998}. In the rest of this letter, we quantify the backaction induced on the condensate due to measurement of the light.

%In deriving Eq.~(\ref{premeasure}) we  have assumed that no light is lost at the mirrors or beam splitters, and that absorption by the atomic sample is likewise negligible, so that the output light fields may be expressed in terms of their coherent amplitudes. We also note that the additional phase rotations due to the propagation of the light cancel out if the interferometer arms are properly balanced.

%The phase is balanced between the two arms at the final beam splitter, thus mixing the ``reference'' beam at a 90 degree phase to the probe beam and ultimately leading to a linear signal in phase shift.  
Assuming the photodectors have unit efficiency, the joint probability of obtaining photon counts $C_+$ and $C_-$ at the first and second photodiodes, respectively, is
\begin{eqnarray}
\label{pureprob}%%%%
 P(C_+,C_-) &=& e^{-|A_0|^2}\frac{(|A_0|^2/2)^{C_++C_-}}{C_+!C_-!}\sum_{|\mathbf j|=n}\frac{n!}{\mathbf j!} \mathbf p^{\mathbf j}\nonumber \\
&\times&(1+\sin\mathbf G\cdot\mathbf j)^{C_+}(1-\sin\mathbf G \cdot \mathbf j)^{C_-}.
\end{eqnarray}
%In that idealized case the atoms will remain in a pure state after the measurement, with
%\be
%\label{postmeasure}%%%%
%|\psi(C_+,C_-)\rangle =\mathcal{N}\sum_{|\mathbf j|=n}\sqrt{\frac{n!}{\mathbf j!}} \mathbf c^{\mathbf j}(1+i e^{-i\mathbf G\cdot\mathbf j})^{C_+}(1-ie^{-i\mathbf G \cdot \mathbf j})^{C_-}|\mathbf j\rangle,
%\ee
%with $\mathcal{N}$ the requisite normalization factor.

In the case $n|\mathbf G|\ll 1$ we find for the mean observed photon count difference the simple expression
\begin{eqnarray}
\label{simplesignal}%%%%
\langle \Delta C\rangle &\approx& |A_0|^2 n\mathbf G \cdot \mathbf p\nonumber \\
&=& |A_0|^2n\left(G_0+\frac{G_+-G_-}{2}\langle F_z\rangle \right . \nonumber \\
&+&\left .\frac{G_+-2G_0+G_-}{2}\langle F_z^2\rangle\right)
\end{eqnarray}
where $\Delta C = C_+ - C_-$ is the difference between photon clicks at the first and second photodiodes, and $\langle F_z\rangle = p_+ - p_-$ and $\langle F_z^2\rangle = p_+ + p_-$ are the single atom spin expectation values. 
%In the previously considered example of $^{87}$Rb the relative sizes of the terms in the signal proportional to 1,  $\langle F_z\rangle$, and  $\langle F_z^2\rangle$ are $6+2\delta : 5-\delta : 1-\delta$, respectively. 

The signal in Eq.~\eqref{simplesignal} can be interpreted as follows. If the atoms are precessing in a magnetic field perpendicular to the light, the phase contrast signal has a large DC component proportional to the number of atoms, an AC signal at the Larmor precession frequency, and a third, smaller signal at twice the Larmor frequency. The amplitude of the respective contributions can be controlled by changing the polarization and/or the detuning of the probe light, thereby changing the relative values of the quantities $\mathbf G$ \cite{carusotto2004}. 
%In our example, it is possible to control the amplitude of each of these contributions by changing $\delta$, confirming its important role as an experimental ``knob.''

In an experiment, the number of condensed atoms is classically uncertain, so the spin state $|\Psi_n\rangle$ is replaced by a density matrix $\hat\rho$. Atom number in an experimental condensate obeys roughly Poissonian statistics:
\be
\hat\rho = e^{-\bar n} \sum_{n=0}^{\infty} \frac{\bar n^n}{n!}|\Psi_n\rangle\langle\Psi_n|
\ee
where $\bar n$ is the mean atom number over many runs and $|\Psi_n\rangle$ is given by Eq.~\eqref{shortinitcond}. Equation \eqref{simplesignal} is modified for this case by making the substitution $n\rightarrow\bar n$.

%{\em Quantum jump operator: }
So far, we have considered the measurement of the condensate state by a pulse of light. To reach a continuous measurement limit, we instead consider coherent light with a photon flux rate per unit time $f$ rather than an amplitude $A_0$, and examine the evolution of the system in a time interval $\delta t$ such that $f\delta t \ll 1$. In this case, the probability of counting no photons at either detector is nearly unity, and any case where multiple photons are counted is negligible. By making the substitution $A_0 \rightarrow \sqrt{f\delta t}$ we can therefore model this continuous observation as a series of short, weak pulses. The free evolution of the atoms will be negligible on this timescale, i.e., between individual photon clicks.

%For an arbitrary initial atomic density matrix $\hat\rho_\mathrm{I}$, the analogues of \eqref{pureprob} and \eqref{postmeasure} to first order in $f\delta t$ are
%\bea
%\label{mixedprob}%%%%
%\nonumber P(0,0) &=& 1-f\delta t,\\
%\nonumber P(1,0) &=& \frac{f\delta t}{2}\sum_{\mathbf j} \rho_{\mathbf j\mathbf j}(1+\sin\mathbf G\cdot\mathbf j),\\
 %P(0,1) &=& \frac{f\delta t}{2}\sum_{\mathbf j} \rho_{\mathbf j\mathbf j}(1-\sin\mathbf G\cdot\mathbf j),\\
%\label{mixedpostmeasure}%%%%
%\nonumber\hat\rho(0,0) &=& (1-f\delta t)\hat\rho_\mathrm{I},\\
%\nonumber\hat\rho(1,0) &=& \frac{f\delta t}{4}\sum_{\mathbf j,\mathbf k} \rho_{\mathbf j\mathbf k} (1+ie^{-i \mathbf G\cdot\mathbf j})|\mathbf j\rangle\langle\mathbf k|(1-ie^{i \mathbf G\cdot\mathbf k}),\\
%\hat\rho(0,1) &=& \frac{f\delta t}{4}\sum_{\mathbf j,\mathbf k} \rho_{\mathbf j\mathbf k} (1-ie^{-i \mathbf G\cdot\mathbf j})
%|\mathbf j\rangle\langle\mathbf k|(1+ie^{i \mathbf G\cdot\mathbf k}).
%\eea
%Here $\rho_{\mathbf j\mathbf k} = \langle \mathbf j|\hat\rho_\mathrm{I}|\mathbf k\rangle$, the sums over vector indices are triple sums with each sub-index separately running from 0 to $\infty$, the final density matrices are conditioned on particular photocount outcomes and have been left unnormalized to reflect the relative probabilities of those outcomes, and all multiphoton counts have probibility of order $(f\delta t)^2$ (or greater).

In this limit, the detection of a photon by either detector (and no photon by the other) has the effect of acting on the density matrix with one of two jump operators:
\bea
\label{jump}
\nonumber\hat\rho(1,0) &=& \hat J_+\hat\rho_\mathrm{I}\hat J_+^\dag,\\
\nonumber\hat\rho(0,1) &=& \hat J_-\hat\rho_\mathrm{I}\hat J_-^\dag,\\
\hat J_\pm &=& \frac{\sqrt{f\delta t}}{2}\left(1\pm ie^{-i\mathbf G\cdot\hat{\mathbf N}}\right).
\eea
The traces of these post-click density matrices represent the probabilities of the respective photodetections. Both probabilities are much less than one; under our assumptions, no click (and hence, no change except a small rescaling of $\hat \rho$) is by far the most likely outcome. In the standard theory of photodetection applied to a cavity with damping $\kappa$ and single mode operator $\hat a$, the jump operator for a photon click is $\sqrt{\kappa\delta t} \hat a$\cite{GardinerCollett}. In the present case, there are two major differences: First, no atoms are removed from the system, and hence the jump commutes with $\hat{\mathbf N}$. Second, as long as $\langle \mathbf G\cdot\hat{\mathbf N}\rangle \ll 1$, a single jump will have only a very small effect on the atomic state. This is to be expected -- a single non-resonant photon passing through our apparatus ought not perturb the atoms much. In the next section, we combine this jump operator with the free spin evolution of the atoms to arrive at a master equation describing the effects of the detection process.

{\em Conditional stochastic master equation:} In accordance with the experimental situation, we consider both the atom-light interaction and the Zeeman interaction induced by the ambient magnetic field,
\be
\label{freeH}
H_\mathrm{Z} = \hbar g_\mathrm{L}\mu_\mathrm{B} \mathbf B \cdot \hat{\mathbf F},
\ee
where $g_\mathrm{L}$ is the Land\'{e} g-factor, $\mu_\mathrm{B}$ is the Bohr magneton, $\mathbf B$ is the applied magnetic field -- eventually the field to be detected -- and $\mathbf F$ is the Schr\"{o}dinger field spin operator. This Hamiltonian could be modified or replaced in more general settings, e.g. by inclusion of the quadratic Zeeman effect. We consider the dynamics of the system over a timescale $\Delta t \gg \delta t$ and an incident flux $f$ such that a natural separation of scales occurs,
\bea
\label{photonscale}
f\Delta t &\sim & \epsilon^{-1},\\
\label{freescale}
g_\mathrm{L}\mu_\mathrm{B} \langle\mathbf B \cdot \hat{\mathbf F}\rangle\Delta t &\sim& \epsilon,\\
\label{jumpscale}
\langle\mathbf G \cdot \hat{\mathbf N}\rangle &\sim& \epsilon^2.
\eea
Crudely speaking, these scales posit a large photon number, small free evolution, and an even smaller effect from a single quantum jump, respectively.

Next, we consider the set of all possible photocounts $\{(C_+,C_-)\}$ and the probabilities of their detection over the interval $\Delta t$. The density matrix will be acted on by jumps $\hat J_\pm$ alternated with unitary evolution due to Eq.~\eqref{freeH}. Because we do not have access to the microscopic details of photon arrival order nor photon arrival times, we must sum (integrate) over the final density matrices for all possible arrival orders (times) consistent with a particular photocount for the interval $\Delta t$. This calculation is aided by the fact that the jump operators will commute with the unitary evolution operators at least to order $\epsilon^2$. The resulting conditional density matrices are normalized by the probability of detection of their particular photocounts. See~\cite{supplemental} for further details or~\cite{Milburn} for a thorough treatment of a similar situation. We use the resulting probability distribution on $\{(C_+,C_-)\}$, and find that, to next to leading order in $\epsilon$,
\be
\label{photonmean}
\langle C_\pm\rangle = V_\pm =\frac{f\Delta t}{2}(1\pm\langle\mathbf G\cdot\hat{\mathbf N}\rangle),
\ee
that is, the mean equals the variance in the counts. The covariance between $C_+$ and $C_-$ vanishes to this order. Also, we note that the means reproduce the signal seen in Eq.~\eqref{simplesignal}. Because the higher moments of the distribution agree with the equivalent Gaussian moments to next to leading order in $\epsilon$, we are able to characterize the photocounts as independent Gaussian random variables,
\be
\label{CVftGN}
C_\pm = \frac{f}{2}(1\pm\langle\mathbf G\cdot\hat{\mathbf N}\rangle)\Delta t + \sqrt{\frac{f}{2}(1\pm\langle\mathbf G\cdot\hat{\mathbf N}\rangle)}\Delta W_\pm,
\ee
where $\Delta W_\pm$ are independent Wiener increments, with
\bea
\label{incrementrules}
\nonumber\langle \Delta W_\pm^2\rangle_\mathrm{E} &=& \Delta t,\\
\langle \Delta W_+\Delta W_-\rangle_\mathrm{E} &=& 0.
\eea
Here the subscript E refers to the ensemble average (that is, over many experimental runs) rather than the usual expectation value. These Wiener increments are Gaussian random variables of zero mean that are functions of time. In the continuum limit, $\Delta \rightarrow d$, these equalities are exact without ensemble averaging. 

The result Eq.~\eqref{CVftGN} is crucial to the derivation of the complete stochastic master equation. We apply the operators $\hat J_\pm$ each $C_\pm$ times to $\rho$, act on it by the free unitary evolution, divide by its trace, and take the continuum limit to obtain~\cite{supplemental},
\be
\label{CSME}
\frac{d\hat\rho}{dt} = \frac{i}{\hbar}[\hat\rho,H^\prime] + \frac{\sqrt{f}}{2}(\mathbf G\cdot\hat{\mathbf N}\hat\rho+\hat\rho\mathbf G\cdot\hat{\mathbf N}-2\langle\mathbf G\cdot\hat{\mathbf N}\rangle\hat\rho)\xi_-(t),
\ee
where the modified Hamiltonian $H^\prime$ is
\be
\label{hprime}
H^\prime = \hbar\left(g_\mathrm{L}\mu_\mathrm{B} \mathbf B \cdot \mathbf F+\frac{1}{2}\mathbf G\cdot\mathbf N\left(f+\sqrt{f}\xi_+(t)\right)\right),
\ee
and  $\xi_\pm(t) = \frac{dV_\pm}{dt}$ are uncorrelated Gaussian white noise functions related to the photocurrents, with $dV_\pm = (dW_+ \pm dW_-)/\sqrt{2}$.

Equations \eqref{CSME} and \eqref{hprime} succinctly encapsulate the full effects of the measurement scheme on the atoms. First, the Hamiltonian is modified by a steady, effective interaction resulting from Eq.~\eqref{compactH}, with the photon number operator replaced by a classical coherent flux. Shot noise in the photon flux leads to an additional random, but unitary, evolution. Finally, the second term in \eqref{CSME} is the non-unitary evolution induced by the measurement of ${\mathbf G}\cdot\hat{\mathbf N}$.

{\em Examples:}
We first examine a very simple application of this master equation to illustrate its salient features in an analytically closed form before moving on to the more experimentally relevant case of the measurement of a magnetic field. Namely, we will assume that there is no external field, $\mathbf B = 0$, and coupling coefficients $G_+ = -G_- = G, G_0 = 0$. This means that the setup will measure only $\hat F_z$, i.e.,
\be
\mathbf G \cdot \hat{\mathbf N} = G\hat F_z.
\ee
and the total Hamiltonian is quantum non-demolition (QND) for the observable $\hat F_z$.

We can derive equations of motion for the observables of the system by multiplying Eq.~\eqref{CSME} by their operators and taking the trace. Because of the large number of atoms involved, we assume that the Gaussian state ansatz will be valid; thus, only equations for the observables, their variances, and their covariances are needed. The chain rule of It\^{o} calculus (resulting from Eq.~\eqref{incrementrules}) must be used when considering the derivatives of the (co)variances, as they are non-linear functions of the other variables. We obtain, for example,
$d{\langle \hat F_z\rangle}/dt = G\sqrt{f}\xi_-(t)v_z$, where $v_z$ is the variance of $\langle\hat F_z\rangle$, and
$\dot{v}_z = -fG^2v_z^2$, which quickly integrates to
\be
v_z(t)=(v_z(0)^{-1}+fG^2t)^{-1}.
\ee
Note that determination of the mean value depends on the experimental record through $\xi_-(t)$, whereas the variance always decreases as the measurement continues. This should be unsurprising -- continuous observation of $\hat F_z$ leads to greater certainty in its value. Over long times, the uncertainty can drop below the standard quantum limit.
\begin{figure}[t]
\begin{center}
\includegraphics[width=0.5\textwidth]{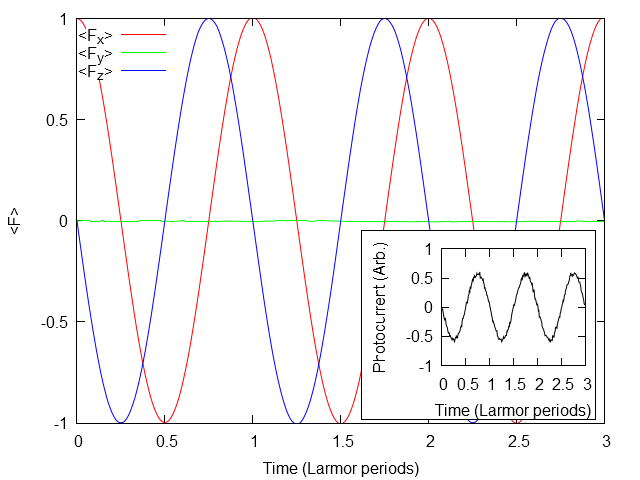}
\end{center}
\caption{\label{figweak}The evolution of the condensate's mean spin along each axis for $I=0.01I_\mathrm{sat}$, normalized by atom number, for a single simulated experimental run. The optical detection of Larmor precession has virtually no effect on the free dynamics. Insert: The normalized photocurrent difference oscillates at the correct frequency. }
\end{figure}
\begin{figure}[b]
\begin{center}
\includegraphics[width=0.5\textwidth]{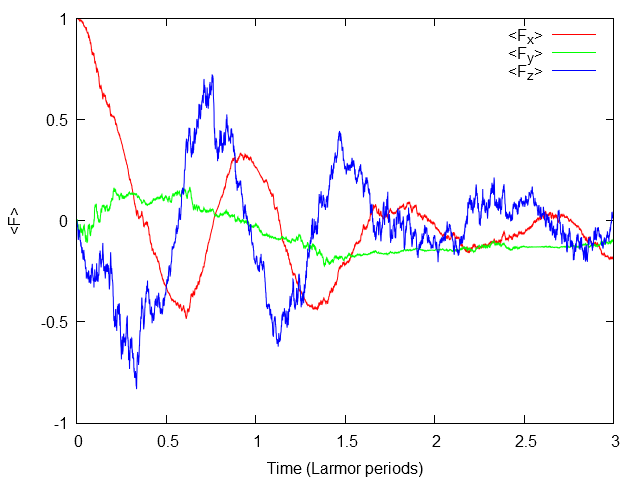}
\end{center}
\caption{\label{figstrong}The evolution of the condensate's mean spin projected along each axis for $I=I_\mathrm{sat}$ for a single run. This stronger measurement causes rapid decay of the condensate's spin.}
\end{figure}
Next, we introduce a magnetic field to be measured, and present the results of a numerical integration of the master equation over a single simulated experimental run. We measure this magnetic field by observing the spin in a transverse direction; the atoms' Larmor precession frequency reveals the field strength.
We assume a condensate of $10^4$ atoms confined within a transverse
spatial extent of 15 $\mu$m in an ambient magnetic field of 1 mG in the $y$-direction. We apply an additional magnetic field $\frac{Gf}{2g_\mathrm{L}\mu_\mathrm{B}}$ in the negative $z$-direction; since $G$ and $f$ are well-controlled experimental parameters this should be possible to a high degree of accuracy. This additional field cancels out the classical portion of the light-matter interaction, though the photon shot noise fluctuations in Eq.~\eqref{hprime} are still present.
The probe is detuned by $2 \pi \times$ 150 MHz below the $F = 1 \rightarrow
F' = 2$ (D1) transition of $^{87}$Rb, and interacts with the
condensate for a total measurement time of 100 ms. The evolution of
the condensate spin is shown for probe intensities of (Fig.~\ref{figweak}) $I = 0.01I_\mathrm{sat}$ and (Fig.~\ref{figstrong}) $I = I_\mathrm{sat}$, where $I_{sat}$ is the saturation
intensity. The measurement strength can best be characterized by the dimensionless ratio $\frac{Gf}{\nu_{\mathrm L}}$, where $\nu_{\mathrm L}$ is the Larmor frequency.
For the parameters listed, this ratio is 0.1 and 10., respectively. In the former case, the free evolution of the atoms is not noticeably perturbed; yet the photocurrent signal unambiguously oscillates at the Larmor frequency. This will allow an accurate determination of the applied field. On the other hand, as the probe intensity is increased and the measurement strength exceeds unity, the backaction-induced stochastic evolution overwhelms the free Larmor precession, resulting in a rapid decay of the transverse magnetization of the condensate in only a few oscillations. This crossover suggests a dynamical phase transition as one moves between the weak and strong measurement regimes.

{\em Summary:}
We have provided a theoretical treatment of the quantum backaction due to the dispersive interaction between a spinor Bose-Einstein condensate and an off-resonant light field. In addition to being the basis for optical magnetometry using a Bose condensate, this interaction has also been shown to be a versatile quantum interface for quantum information processing and state engineering \cite{hammerer2010}. Straightforward additions to our model include a description of spatially inhomogeneous spin textures in the condensate, stroboscopic optical measurements of the condensate and the detection of time-varying magnetic fields. In addition to understanding the potential sensitivity of
condensate-based magnetic field sensing, our formalism can also be
applied to quantum-limited nondestructive imaging of magnetic textures
in spinor condensates and the creation of novel many-body states via
quantum non-demolition (QND) measurement \cite{mekhov2012}.

%We have presented here a theoretical treatment of a magnetometer comprised of the atoms in a Bose-Einstein condensate and probed by light. As experimental techniques progress, they will inevitably reach the levels of sensitivity where a fully quantum description of the system is necessary to understand its potential sensitivity. This work provides a general guide to the development of accurate modeling for such experiments.
%
%Many additional features can be added with little difficulty to this model. We anticipate that in the diversity of experimentally feasible situations it will be necessary to take into account motional states of the atoms, the quadratic terms in the free Hamiltonian, spin texturing and domains, different light polarizations, pulsed rather than continuous measurements, time-varying magnetic fields, spatial resolution, optical absorption, alternate parameter regimes where the system may behave quite differently, and of course several more effects.

{\em Acknowledgements}: This work was supported by the DARPA QuASAR program through a grant from AFOSR and the DARPA ORCHID program through a grant from ARO, the US Army Research Office, and by NSF. M. V. acknowledges support from the Alfred P. Sloan Foundation. The authors would also like to thank Carlo Samson and Chandra Raman of the Georgia Institute of Technology for useful input on additional experimental 
considerations.
\bibliography{magnetometrybib}
\end{document}

\title{Supplemental material for``Quantum back-action in spinor condensate magnetometry''} 
\author{S. K. Steinke}
\affiliation{B2 Institute, Department of Physics and College of Optical Sciences\\The University of Arizona, Tucson, Arizona, 85721.}

\author{S. Singh}
\affiliation{ITAMP, Harvard-Smithsonian Center for Astrophysics, Cambridge, Massachusetts 02138.}

\author{P. Meystre}
\affiliation{B2 Institute, Department of Physics and College of Optical Sciences\\The University of Arizona, Tucson, Arizona, 85721.}

\author{K. C. Schwab}
\affiliation{Applied Physics, California Institute of Technology, MC 128-95, Pasadena, California 91125.}

\author{M. Vengalattore}
\affiliation{Laboratory of Atomic and Solid State Physics, Cornell University, Ithaca, New York, 14853.}

\date{\today}
\maketitle
\section{Details of a specific light-matter interaction Hamiltonian}
As a concrete example of a light-matter Hamiltonian of the type in the main paper, we examine the case of circularly polarized light incident on a condensate of $^{87}$Rb atoms. We consider only contributions from the D$_1$ virtual transitions between the ground state $S_{1/2}$ and excited $P_{1/2}$ manifolds but add the effects of the virtual transitions from the $F=1$ to both the $F^\prime = 1$ and $F^\prime = 2$ submanifolds. We find in the rotating wave approximation and after adiabatically eliminating the excited atomic states
\be
H_{I}=\frac{\hbar \Omega^2}{48 \Delta_2}\left [6\hat N_++(3+\delta)\hat N_0 +(1+\delta)\hat N_-\right ]\hat b^\dag \hat b.
\ee
Here the spin basis is in the direction of light propagation, $\Omega$ is the single photon Rabi frequency, $\Delta_2$ is the detuning (possibly negative) for the $F=1\rightarrow F'=2$ transition, $\Delta_1=\Delta_2+814.5$ MHz is the $F=1\rightarrow F'=1$ detuning, and $\delta$ is the experimentally adjustable ratio given by
\be
\delta = \frac{\Delta_2}{\Delta_1}.
\ee
This parameter adds a powerful element of tunability to the observable measured by the phase shift. 

\section{Derivation of the Conditional Stochastic Master Equation}
As mentioned in the text, the unitary evolution operator commutes with the jump operators to at least order $\epsilon^2$, so we can write down the (unnormalized) conditional density matrix after $\Delta t$ has passed as
\be
\label{conditionalfinitestep}
\hat\rho(t+\Delta t;C_+,C_-) = U(\Delta t)J_+^{C_+}J_-^{C_-}\hat\rho(t)J_-^{\dag C_-}J_+^{\dag C_+}U^\dag(\Delta t),
\ee
where $U$ is the free unitary evolution, and $C_\pm$ is given by
\be
\label{CVftGN}
C_\pm = \frac{f}{2}(1\pm\langle\mathbf G\cdot\hat{\mathbf N}\rangle)\Delta t + \sqrt{\frac{f}{2}(1\pm\langle\mathbf G\cdot\hat{\mathbf N}\rangle)}\Delta W_\pm.
\ee
In the present situation where the condensate operates as a magnetometer this evolution is given by
\be
U(\Delta t) = e^{-i g_\mathrm{L}\mu_\mathrm{B} \mathbf B \cdot \hat{\mathbf F}\Delta t}.
\ee

Constant factors can be removed from Eq.~\eqref{conditionalfinitestep}, as $\hat\rho$ needs to be normalized later, anyway. We therefore remove factors of $(1/2)(1\pm i)\sqrt{f\delta t}$ from $\hat J_\pm$, making the operators unity to zeroth order in $\epsilon$ and simplifying the computation (to next to leading order) of
\be
\label{JC}
J_+^{C_+}J_-^{C_-} = 1-\frac{i}{2}\mathbf G\cdot\hat{\mathbf N}(f\Delta t - \sqrt{f}\Delta V_+)+\frac{1}{2}\sqrt{f}\Delta V_-,
\ee
where 
\be
\Delta V\pm = \frac{\Delta W_+ \pm \Delta W_-}{\sqrt{2}}
\ee
are also independent (from each other) Wiener increments. After substituting Eq.~\eqref{JC} into Eq.~\eqref{conditionalfinitestep}, we obtain the trace of $\hat\rho(t+\Delta t;C_+,C_-)$,
\be
\label{tracerhoc}
\mathrm{tr}\hat\rho(t+\Delta t;C_+,C_-) = 1+\langle\mathbf G\cdot\mathbf N\rangle\sqrt{f}\Delta V_-+\mathcal{O}(\epsilon^2).
\ee
We then divide Eq.~\eqref{conditionalfinitestep} by Eq.~ \eqref{tracerhoc} and  take the continuum limit to obtain the final conditional stochastic master equation and modified Hamiltonian listed in the text,
\be
\label{CSME}
\frac{d\hat\rho}{dt} = \frac{i}{\hbar}[\hat\rho,H^\prime] + \frac{\sqrt{f}}{2}(\mathbf G\cdot\hat{\mathbf N}\hat\rho+\hat\rho\mathbf G\cdot\hat{\mathbf N}-2\langle\mathbf G\cdot\hat{\mathbf N}\rangle\hat\rho)\xi_-(t),
\ee
\be
\label{hprime}
H^\prime = \hbar\left(g_\mathrm{L}\mu_\mathrm{B} \mathbf B \cdot \mathbf F+\frac{1}{2}\mathbf G\cdot\mathbf N\left(f+\sqrt{f}\xi_+(t)\right)\right),
\ee
where $\xi_\pm(t) = \frac{dV_\pm}{dt}$ are uncorrelated Gaussian white noise functions, $dV_\pm = (dW_+ \pm dW_-)/\sqrt{2}$
.